\documentclass[fleqn,usenatbib]{mnras}

\usepackage{newtxtext,newtxmath}

\usepackage[T1]{fontenc}

\DeclareRobustCommand{\VAN}[3]{#2}
\let\VANthebibliography\thebibliography
\def\thebibliography{\DeclareRobustCommand{\VAN}[3]{##3}\VANthebibliography}
\usepackage[dvipsnames]{xcolor}
\usepackage{soul}

\newcommand{\red}[1]{{ \textbf{#1} }}

\usepackage{subfigure}

\usepackage{graphicx}	
\usepackage{amsmath}

\title[Magnetic equilibria in stars]{Stability of axially symmetric magnetic fields in stars}

\author[Becerra et al.]{
Laura Becerra,$^{1}$\thanks{E-mail: lbecerra@astro.puc.cl}
Andreas Reisenegger,$^{2}$
Juan Alejandro Valdivia,$^{3,4}$
Mikhail Gusakov$^{5}$
\\
$^{1}$Instituto de Astrof\'isica, Facultad de F{\'i}sica, Pontificia Universidad Cat\'olica de Chile, Av. Vicu\~na Mackenna 4860, Macul, Santiago, Chile\\
    $^{2}$Departamento de F\'{\i}sica, Facultad de Ciencias B\'asicas, Universidad Metropolitana de Ciencias de la Educaci\'on, Av. Jos\'e Pedro Alessandri 774,\\ \~Nu\~noa, Santiago, Chile\\
$^{3}$Departamento de F\'{\i}sica, Facultad de Ciencias,
Universidad de Chile, Las Palmeras 3425, \~Nu\~noa, Santiago, Chile\\
$^{4}$Centro para el Desarrollo de la Nanociencia y
Nanotecnolog\'\i a, CEDENNA, Santiago, Chile\\
$^{5}$Ioffe Institute, 26 Politekhnicheskaya Street, St. Petersburg 194021, Russia
}

\date{Accepted XXX. Received YYY; in original form ZZZ}

\pubyear{2022}

\defcitealias{paperI}{Paper~I}
\begin{document}
\label{firstpage}
\pagerange{\pageref{firstpage}--\pageref{lastpage}}
\maketitle

\begin{abstract}
The magnetic fields observed in Ap-stars, white dwarfs, and neutron stars are known to be stable for long times. However, the physical conditions inside the stellar interiors that allow these states are still a matter of research.  
It has been formally demonstrated that both purely toroidal and purely poloidal magnetic fields develop instabilities at some point in the star.  On the other hand, numerical simulations have proved the stability of roughly axisymmetric magnetic field configurations inside stably stratified stars. These configurations consist of mutually stabilizing toroidal and poloidal components in a twisted torus shape. Previous studies have proposed rough upper and lower bounds on the ratio of the magnetic energy in the toroidal and poloidal components of the magnetic field. With the purpose of mapping out the parameter space under which such configurations remain stable, we used the Pencil Code to perform 3D magnetohydrodynamic simulations of the evolution of the magnetic field in non-rotating, non-degenerate stars in which viscosity is the only dissipation mechanism, both for stars with a uniform (barotropic) and radially increasing (stably stratified) specific entropy. Furthermore, we considered different conditions regarding the degree of stable stratification and the magnetic energy in each component, roughly confirming the previously suggested stability boundaries for the magnetic field. 

\end{abstract}




\section{Introduction}

The magnetic fields observed in stars can be grouped into two general types based on their phenomenology. 
The first group is formed by the magnetic fields observed in low-mass non-degenerate stars. These have a complex structure,  are variable on time scales that go from weeks to decades, and are probably produced by a contemporary dynamo in the star's convective envelopes \citep{2009IAUS..259..339R}. In the second group are the magnetic fields detected in some intermediate and high-mass main-sequence stars, white dwarfs, and neutron stars, which are long-lived and globally organized. The partly (or completely) stably stratified interiors of these stars are not compatible with a dynamo origin for their magnetic field. These fields are  slowly varying or stable, likely having been set up in an earlier stage of the evolution of these stars \citep{2009ARA&A..47..333D,2019MNRAS.483.3127S}.

It has yet to be fully understood what conditions need to be satisfied by the latter group of stars and their  magnetic field configuration so that they are able to survive over the star's lifetime.  
Since \cite{1953ApJ...118..116C}, many works have been devoted to building magnetic field configurations by solving the hydromagnetic equilibrium equation numerically or analytically and to prove the stability of these configurations over dynamical timescales. 
\citet{1973MNRAS.161..365T} found general conditions for the stability of purely toroidal fields\footnote{Within the Cowling approximation (that is, neglecting perturbations of the gravitational potential; \citealt{1941MNRAS.101..367C}), these conditions are necessary and sufficient for stability. Relaxing this approximation, the conditions remain necessary, but not sufficient.} and used them to show that any such field having a finite current density along the axis will be unstable in this region. The application of these conditions to more general field configurations \citep{Goossens+1981,2013MNRAS.433.2445A} proved that all toroidal fields are unstable.
Similarly, purely poloidal magnetic fields were also found to be subject to dynamical instabilities \citep{1973MNRAS.163...77M,1973MNRAS.162..339W,1977ApJ...215..302F}, suggesting that a combination of poloidal and toroidal components is required for the stability of axially symmetric magnetic fields \citep{1956ApJ...123..498P}.

Later, the numerical simulations of \citet{2006A&A...450.1077B} \cite[see also][]{2004Natur...431..819B} found that roughly axisymmetric stable equilibria, consisting of both toroidal and poloidal components of comparable strength in a twisted-torus shape, formed spontaneously from initially random magnetic field configurations in stably stratified stars. In \citet{paperI} (hereafter Paper~I), we showed that the formation of these axially symmetric magnetic fields is strongly influenced by the use of a high-order diffusion scheme introduced for numerical stability purposes, particularly by the presence of a substantial magnetic diffusivity. Based on our simulations, we argued that, under more realistic physical conditions, the magnetic field configuration is likely to evolve to a stable, but non-axially symmetric state. However, due to their simplicity, the axially symmetric field configurations are the most widely studied among the community, and observations seem to support the existence of such equilibrium configurations in at least some stars \citep{2007A&A...475.1053A}. 

In such configurations, the poloidal and toroidal magnetic field components need to stabilize each other. For the stabilization of a poloidal field, the toroidal field must be at least of a similar strength, as first suggested by \citet{1973MNRAS.163...77M} and later confirmed by \citet{2009MNRAS.397..763B}, whose simulations yielded the approximate condition that the energy in the poloidal component, $E_{\rm pol}$, can be at most $\sim 80\%$ of the total magnetic energy $E_{\rm mag}$, for the field to be stable.

For the stability of the toroidal field, it is not enough to invoke the presence of a poloidal component, but it is also necessary for the stellar matter to be stably stratified, with a positive squared Brunt-V\"ais\"al\"a (or buoyancy) frequency, 
\begin{equation}\label{eq:buoyancy}
N^2\equiv\left({1\over\gamma}-{1\over\Gamma}\right){\rho g^2\over p}>0,
\end{equation}
where $\rho$, $p$, and $g$ are the local values of the fluid density, pressure, and acceleration of gravity, respectively. $\Gamma$ is the adiabatic index,
\begin{equation}\label{eq:adia_index}
    \Gamma= \left(\frac{\partial {\rm ln}\, p}{\partial {\rm ln}\, \rho}\right)_X,
\end{equation}
where $X$ represents one or more conserved quantities characterizing adiabatic perturbations, such as specific entropy in the case of main sequence stars and white dwarfs, and relative particle abundances in the case of neutron stars \citep{2009A&A...499..557R}, while $\gamma$ is the conventional polytropic index:
\begin{equation}\label{eq:poly_index}
    \gamma = \frac{d\, {\rm ln}\, p}{d\, {\rm ln}\, \rho}\, ,
\end{equation}
characterizing the equilibrium inside the star. In the astrophysical examples mentioned above, the typical buoyancy frequency is much larger than the Alfv\'en frequency, $\omega_A=B/(\sqrt{\mu_0\rho} R_s)$, where $B$ is the magnetic field strength, $R_s$ is the stellar radius and $\mu_0$ is the magnetic vacuum permeability. In this case, the stable stratification strongly suppresses radial motions. In particular, instabilities of the toroidal field require displacements with a non-zero radial component, but this must be much smaller than the horizontal components if the star is strongly stratified \citep{2013MNRAS.433.2445A}, and this helps the poloidal component to stabilize such motions.  \citet{2009MNRAS.397..763B} used heuristic arguments to write the condition for the stabilization of the toroidal component as $E_{\rm pol}/E_{\rm mag}\gtrsim aE_{\rm mag}/|E_{\rm grav}|$, where $E_{\rm grav}$ is the star's gravitational energy and $a$ is a dimensionless constant that is inversely proportional to the difference $\Gamma-\gamma$, and which he found numerically to be $\sim 10$ for the case of main-sequence stars, for which $\Gamma\approx 5/3$ and $\gamma\approx 4/3$. Through a mostly analytic calculation, \citet{2013MNRAS.433.2445A} suggested $a\approx 1.8/(\Gamma/\gamma-1)$, in rough agreement with Braithwaite's result.

The two conditions for the mutual stabilization of the toroidal and poloidal field components  discussed above can be written together as\footnote{Here it is assumed that $N\gg\omega_A$, which implies that the term on the right-hand side is large, as is the case in the stars of interest and in all the simulations used to obtain the stability limits.}
\begin{equation}\label{eq:stb_limit}
    0.25 \lesssim \frac{E_{\rm tor}}{E_{\rm pol}} \lesssim 0.5 \sqrt{ \left(\frac{\Gamma}{\gamma} - 1 \right)\frac{ |E_{\rm grav}|}{E_{\rm pol}} },
\end{equation}
showing the two essential ingredients for the stability of the magnetic field in the stellar interior: the relative strengths of the toroidal and poloidal components of the magnetic field, and the star's stable stratification ($\Gamma>\gamma$). The importance of the latter effect was confirmed by simulations that failed to find any stable configuration in  barotropic (non-stratified, $\Gamma=\gamma$) stars \citep{2012MNRAS.424..482L,2015MNRAS.447.1213M}.

We note that, even disregarding the issue of stability, there is an important difference between hydromagnetic equilibria in barotropic stars (with $N=0$) and those with strong stable stratification ($N^2\gg\omega_A^2$). In the latter, pressure and density can be treated as independent variables, so the fluid has two scalar degrees of freedom, and the only condition for an axially symmetric magnetic field to correspond to a possible equilibrium state is that the toroidal component of the Lorentz force vanishes, leading to the condition $\beta=\beta(\alpha)$ for the scalar functions introduced in Equation~(\ref{eq:B_axial}). In the former, there is a one-to-one relation between pressure and density, so the fluid has only one scalar degree of freedom, and the functions $\alpha$ and $\beta$ are additionally required to satisfy the Grad-Shafranov equation \citep{grad1958,1958JETP....6..545S}. This is a highly nonlinear equation, whose general solution space has not yet been characterized. Various authors have explored numerical solutions, generally finding only configurations with $E_{\rm tor}/E_{\rm pol}<0.1$ \citep[see for example][among others]{Yoshida+2006,Lander+2009,Ciolfi+2009,Fujisawa+2012, Armaza+2015}, which was explained through an approximate analytical argument in \citet{Armaza+2015}. If the magnetic field configuration is totally confined inside the star, the $E_{\rm tor}/E_{\rm pol}$ fraction could become larger \citep{Haskell+2008, 2010A&A...517A..58D}. However, \cite{2012MNRAS.424..482L} studied the stability of these configurations, finding that they were all unstable in barotropic stars. 

Equation~(\ref{eq:stb_limit}) also has important astrophysical consequences. For example, since $E_{\rm mag}\ll|E_{\rm grav}|$, a star can store a (internal) toroidal component much stronger than the poloidal field measured on its surface. If this is the case,  the anisotropic pressure of the internal magnetic field deforms the star into a prolate shape  \citep{1953ApJ...118..116C}.  This could cause the  precession \citep{2003MNRAS.341.1020W} inferred in some pulsars, such as PSR B1828-11 \citep{2000Natur.406..484S,2017MNRAS.467..164A}, which has also been suggested to be relevant for fast radio bursts (FRBs) from magnetars  \citep{2022ApJ...928...53W}. It would also lead to the emission of continuous gravitational waves by rotating neutron stars \citep{2009MNRAS.398.1869D}, contributing to their  spin-down \citep{2002PhRvD..66h4025C, 2006MNRAS.365..653A}. On the other hand, a strong internal toroidal field could provide an energy source for the violent activity of magnetars \citep{1996ApJ...473..322T,2008A&ARv..15..225M} and the large pulsed fraction in the X-ray emission of central compact objects \citep{2012ApJ...748..148S,2012MNRAS.425.2487V}.

In this paper, we address the problem of the stability of axially symmetric magnetic fields in stably stratified stars by simulating the dynamical evolution of many different configurations, in order to verify  the accuracy of Equation~(\ref{eq:stb_limit}). For this, we use the {\sc Pencil Code} \citep{2021JOSS....6.2807P}, a high-order finite-difference code for compressible hydrodynamic flows with magnetic fields. For all the simulations presented in this paper, the only dissipative mechanism considered is ordinary shear viscosity, while all hyper-diffusion coefficients are taken to vanish. As argued in \citetalias{paperI}, this approach is expected to most closely mimic the physics of the real stars. Section~\ref{sec:Eqs_code} gives a brief description of the numerical setup and initial conditions, and section~\ref{sec:stbl_anal} is devoted to testing the stability of different axially symmetric magnetic field configurations as a function of the ratio of magnetic to gravitational energy, the fraction of the magnetic energy contained in the poloidal and toroidal components, and the stable stratification of the stellar matter.  Our conclusions are presented in section~\ref{sec:conclusion}.

%
%

\section{Initial setup}\label{sec:Eqs_code}

We test the stability of axially symmetric magnetic field configurations by simulating their dynamical evolution inside the star with the {\sc Pencil Code}\footnote{ \texttt{https://github.com/pencil-code/}} \citep{2021JOSS....6.2807P}, which uses sixth-order centered spatial derivatives and a third-order Runge-Kutta time-stepping scheme to solve the magneto-hydrodynamics (MHD) equations (given in Appendix~\ref{ap:equations}).   
All the simulations are performed in a  cubic box of side $L_{\rm box}=4.5 R_{\rm s}$, 
centered on the star, with a Cartesian grid with $128^3$ equally spaced points and periodic boundary conditions. 

The fluid inside the simulation box obeys the ideal gas equations of state with an adiabatic index  $\Gamma=5/3$ (see Equation~[\ref{eq:adia_index}]), with specific entropy as the stabilizing quantity, $X=s$. As done in \citetalias{paperI}, to prepare the initial condition for the simulations, we first built a non-rotating, unmagnetized spherical star model. Thus, we solved the (non-magnetic) hydrostatic equilibrium equations, adopting, inside the star,  a polytropic relation between the gas pressure and density: $p=K \rho^{\gamma}$, where $K$ and $\gamma$ are constants. When $\gamma<\Gamma$, the star is stably stratified (the specific entropy is an increasing function of the radius), while for $\gamma=\Gamma$, it is barotropic (no entropy gradient). For the initial setup of the barotropic case ($\Gamma=\gamma=5/3$) and the slightly stratified case ($\Gamma=5/3$ and $\gamma=1.61$), it was not possible to apply the polytropic relation to the whole stellar interior, because this made the density near the surface very low, causing problems to the numerical scheme. Thus, for these cases, the specific entropy was taken to increase for $r>(0.97-0.98)R_s$, creating a thin, stably stratified surface layer, as seen in Figure~1 of \citetalias{paperI}.

Outside the star, we placed a low-density atmosphere with a uniform temperature and a low electrical conductivity, to make the magnetic field quickly relax to a potential field. Inside the star, the electrical conductivity was assumed to be very large.  So, we set the magnetic diffusivity equal to zero inside the star and large and constant outside it \citepalias[for a complete account of the initial setup, we refer the reader to ][]{paperI}. 

For this spherically symmetric equilibrium configuration, we calculated the gravitational potential, which was kept constant in all our simulations. We note that by doing this, we apply the Cowling approximation \citep{1945MNRAS.105..166C} in two different ways. First, we neglect the perturbation of the gravitational potential produced by the presence of the initial magnetic field, since the magnetic equilibria of the star can be seen as a small perturbation of its non-magnetic spherical background. And second, we also neglect the changes in the gravitational potential along the simulations, due to the evolution of the magnetic field (The latter is the ``standard'' Cowling approximation used by \citet{1973MNRAS.161..365T} and others.)

Once this was done, we introduced the desired initial magnetic field and established the initial magnetic equilibrium by running the code for a few sound crossing timescales while freezing the evolution of the magnetic field, allowing the hydrodynamical forces to balance the Lorentz force. The latter was only done for the stably stratified models, since the initial magnetic field configurations we chose do not satisfy the Grad-Shafranov equation, so they do not correspond to equilibria for barotropic stars (see the discussion in the Introduction). Thus, for this case, it was not attempted to make the fluid relax to an equilibrium before starting the simulation.

Finally, to accelerate the appearance of eventual instabilities, if not specified otherwise,
a small random perturbation was added to the density field, and the simulation was continued, now using the full set of equations described in Appendix~\ref{ap:equations}, so the magnetic field was allowed to evolve.  

In all the simulations, the viscosity coefficient is  $\nu=8.24\times 10^{-4} R_s/\tau_s$. We note that our results are independent of  the dissipation mechanism for the kinetic or the magnetic energy, which we verified by also running simulations with the hyper-diffusion scheme studied in \citetalias{paperI}, and obtaining the same results for the stability of each magnetic field configuration considered. We refer the reader to \citetalias{paperI} for more details of the initial setup.  

\subsection{Magnetic field configuration}\label{sec:setup}
%
\begin{figure}
    \centering
    \subfigure[Field I.]{\includegraphics[width = 0.48\columnwidth]{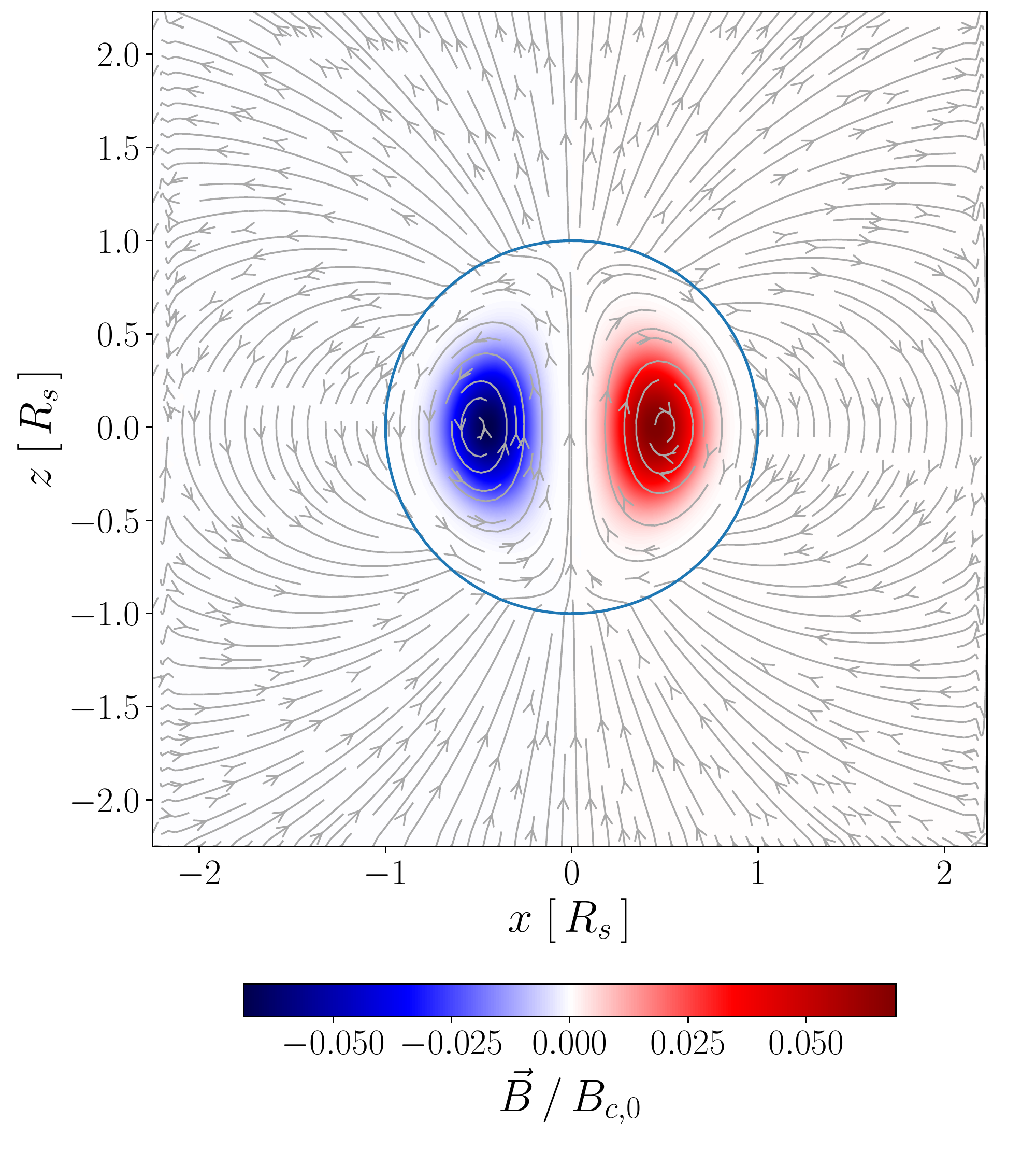} }
    \subfigure[Field II.]{\includegraphics[width = 0.48\columnwidth]{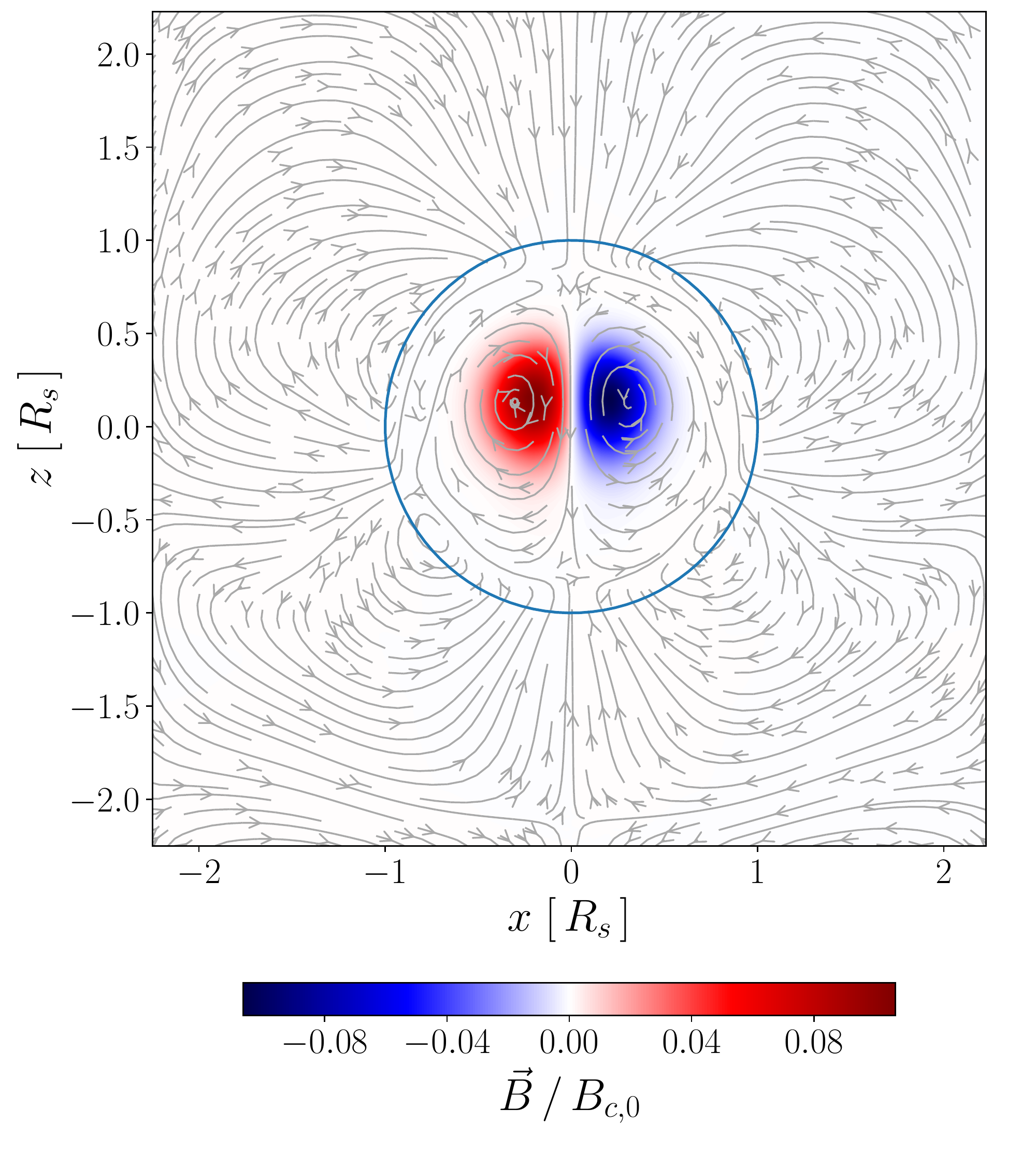} }
    \caption{View of the two choices of axially symmetric magnetic field configuration in the star's meridional plane, with the symmetry axis in the vertical direction: (a) Field I: analytic expression from \citet{2013MNRAS.433.2445A}; (b) Field II: obtained numerically from a random initial field (\citetalias{paperI}; see text for details). }
    \label{fig:Field_axiall}
\end{figure}

In order to verify the validity of Equation~(\ref{eq:stb_limit}), we follow the evolution of ordered axially symmetric initial magnetic field configurations in the stellar interior, under different initial conditions for the star's stable stratification (or its absence), total magnetic energy, ratio between the energies in the poloidal and toroidal components of the magnetic field, and two different functional forms for the magnetic field structure, ``Field I'' and ``Field II'', shown in Figure~\ref{fig:Field_axiall}.   

For Field I, we use the configuration built analytically in \cite{2013MNRAS.433.2445A}.  In spherical coordinates $(r,\theta, \phi)$, an axially symmetric magnetic field can be written as the sum of a poloidal and a toroidal component: 
\begin{equation}\label{eq:B_axial}
    \vec{B} = \vec{B}_{\rm pol} + \vec{B}_{\rm tor} =   a_{\rm pol}\vec{\nabla}\alpha(r, \theta) \times \vec{\nabla}\phi +   a_{\rm tor}\beta(r, \theta)  \vec{\nabla}\phi  
\end{equation}
\citep{1956PNAS...42....5C}, where $\alpha(r,\theta)$ and $\beta(r,\theta)$ are scalar functions further discussed below, $\vec{\nabla}\phi = \hat{\phi} / (r\sin\theta)$, and  $a_{\rm pol}$ and $a_{\rm tor}$ are adjustable coefficients to set the strengths of the poloidal and toroidal components.  
In equilibrium, the azimuthal component of the Lorentz force must vanish, implying $\beta= \beta(\alpha)$. For the poloidal component of the magnetic field, \cite{2013MNRAS.433.2445A} adopted a simple form that makes the external field a pure vacuum dipole:
\begin{equation}\label{eq:alpha}
    \alpha(x,\theta) = f(x)\sin^2\theta \, ,
\end{equation}
with $x=r/R_s$ and
\begin{equation}\label{eq:f_x}
   f(x)= \left\{ 
     \begin{aligned}
f_2x^2 + f_4x^4 + f_6x^6 + f_8x^8 & \quad&
\mathrm{for}&\quad
x \leq 1, \\
 x^{-1} &\quad&
\mathrm{for}&\quad
  x>1.
\end{aligned} 
\right.
\end{equation}
The coefficients $f_i$ are set by applying boundary conditions at the stellar surface (continuity of the  magnetic field and the current density), which yield $f_2 = \frac{35}{8}-f_8$, $f_4 = -\frac{21}{4} + 3f_8$, $f_6=\frac{15}{8}-3f_8$, with $f_8$ as a free parameter. The function $\beta$ must be chosen so that the toroidal field is non-zero only on poloidal field lines that close within the star.
In \cite{2013MNRAS.433.2445A}, it is taken to be 
\begin{equation}\label{eq:beta}
  \beta(\alpha) = \left\{ 
     \begin{aligned}
 \left(\alpha - 1 \right)^2  & \quad&
\mathrm{for}&\quad
\alpha \geq 1,\\
 0 &\quad&
\mathrm{for}&\quad
  \alpha < 1.
\end{aligned} 
\right.
\end{equation}
We choose the free parameter $f_8=-1000$, so that the torus containing the toroidal field lines extends over a large fraction of the stellar volume.

For Field II, we use a magnetic equilibrium obtained  numerically from the evolution of a random initial configuration. Specifically, we take the  magnetic field from Model V of \citetalias{paperI}  at $t=240~\tau_{A,0}$, from which we build an axially symmetric configuration by taking the azimuthal average of each spherical component of the magnetic field ($B_r, B_\theta, B_\phi$) around the  magnetic axis, $\vec{M}$, defined as the direction that minimizes the asymmetry parameter \citepalias[see also ][]{paperI}:
\begin{equation}\label{eq:axial_param}
\mathcal{A}
=\min_{\theta,\phi}\frac{\int_{\rm star} |\vec{B}-\vec{B}^{\rm axial}_{\theta,\phi} |^2 \,dV }{\int_{\rm star} |\vec{B} |^2\, dV } .
\end{equation}
Here, $\vec{B}^{\rm axial}_{\theta,\phi}$ is built by taking the azimuthal average of the magnetic field components in spherical coordinates around a certain axis oriented in the direction given by the angles $\theta$ and $\phi$.

We define the Alfv\'en speed as
\begin{equation}
    v_A\equiv \frac{B_{\rm rms} }{\sqrt{\mu_0\rho_{\rm rms}}},
\end{equation}
and the Alfv\'en and sound crossing time scales as
\begin{equation}
\tau_{A} \equiv \frac{R_s}{v_A} \quad {\rm and} \quad \tau_s\equiv \frac{R_s}{c_{\rm s,rms}},
\end{equation}
respectively, where $c_s$ is the sound speed, and we use  $A_{\rm rms}$ to denote the root-mean-square volume averaged over the simulation box of any quantity $A$.

\section{Stability conditions: Numerical results}\label{sec:stbl_anal}

\subsection{Dependence on the 
poloidal and toroidal magnetic energy}

%
\begin{figure}
    \centering
   \subfigure[]{\includegraphics[width=0.48\columnwidth]{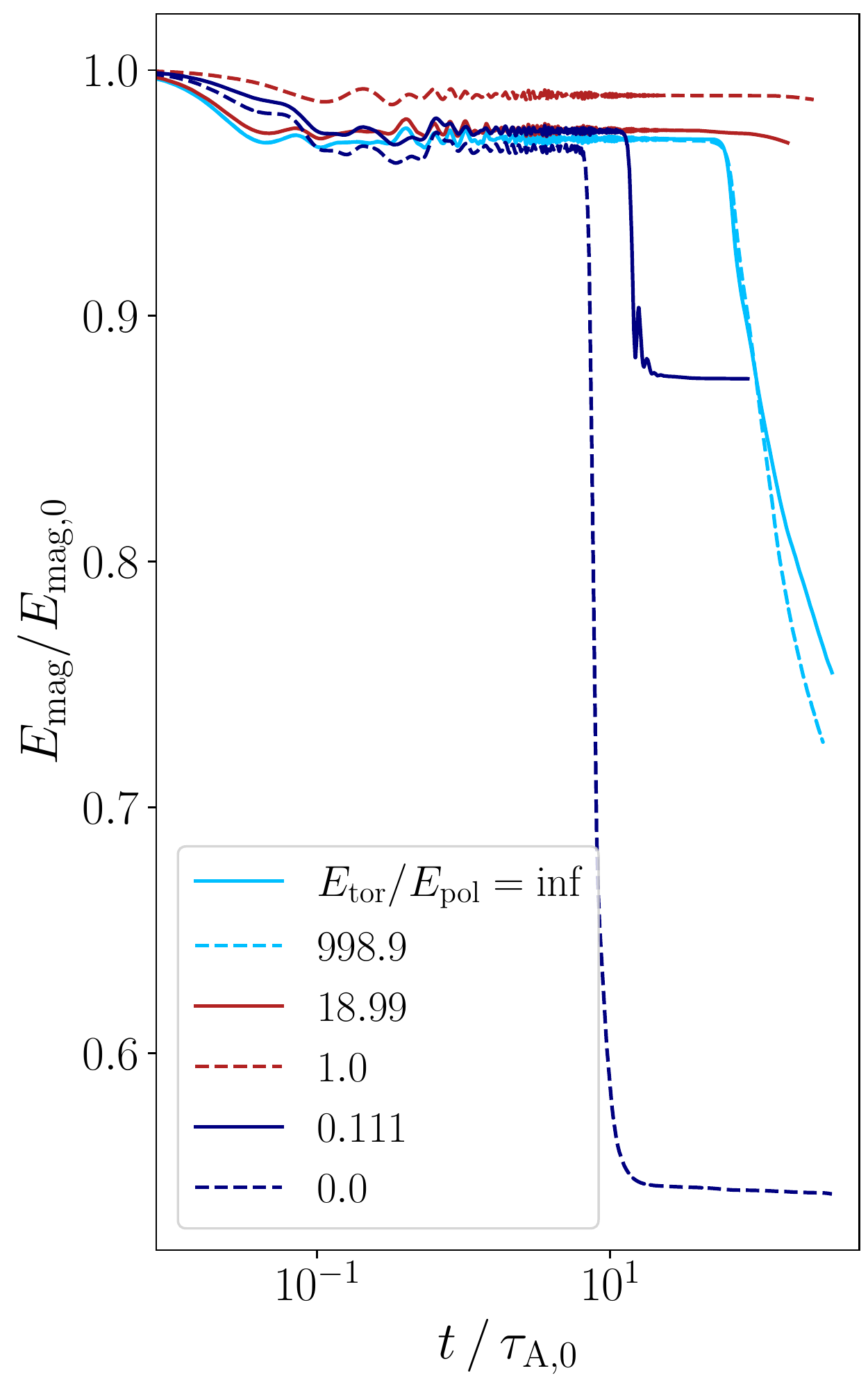}}
   \subfigure[]{\includegraphics[width=0.48\columnwidth]{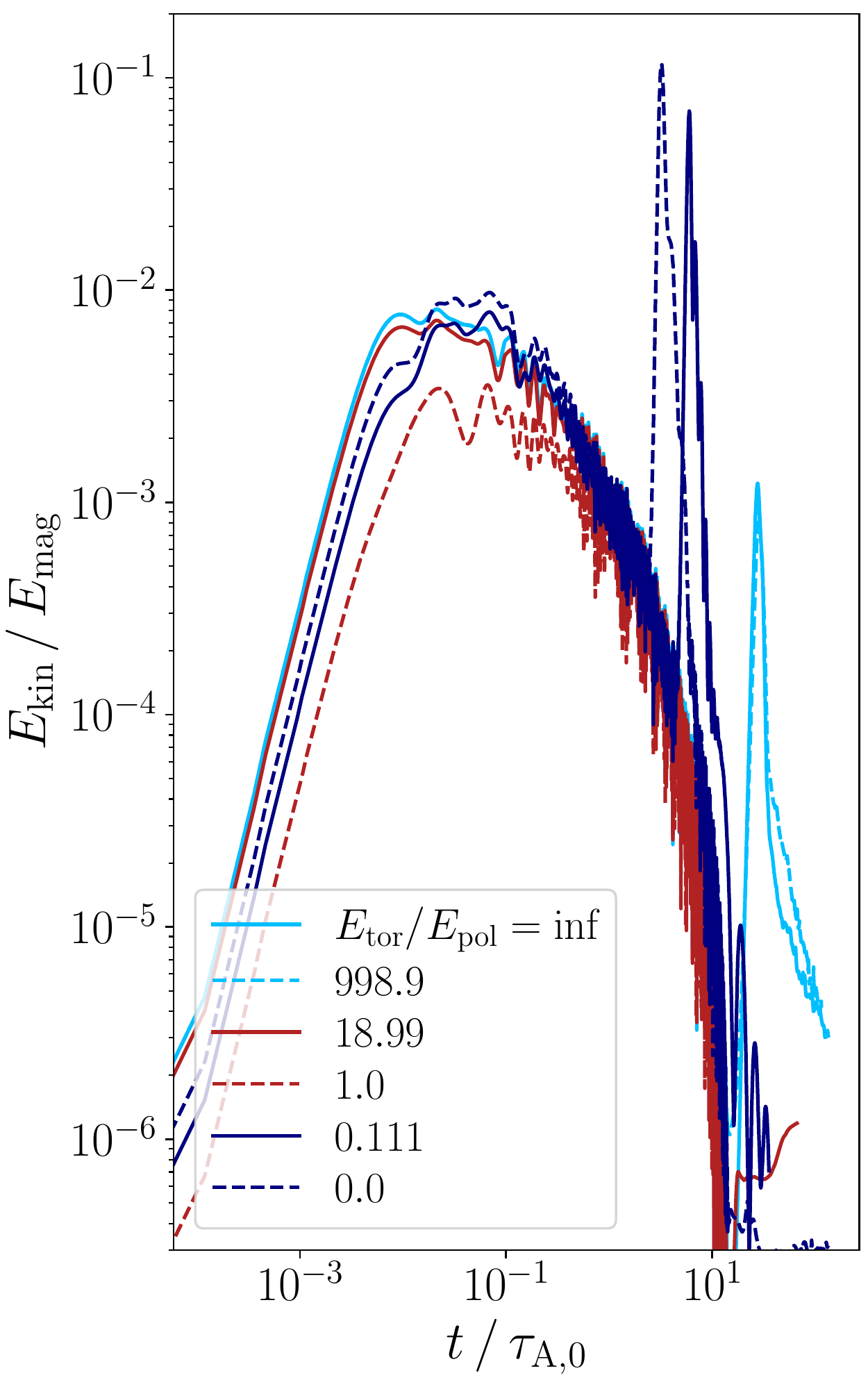}}
    \caption{Time evolution of (a) the total magnetic energy (normalized to its initial value) and (b) the ratio between total kinetic energy and magnetic energy for the initial magnetic field configuration labeled as `Field I' and different initial values of $E_{\rm tor}/E_{\rm pol}$. For all cases, the total initial magnetic energy is $E_{\rm mag}= 2.8\times 10^{-3} |E_{\rm grav}|$.}
    \label{fig:Emag_ArguinE10}
\end{figure}
\begin{figure*}
    \centering
\includegraphics[width=0.99\textwidth]{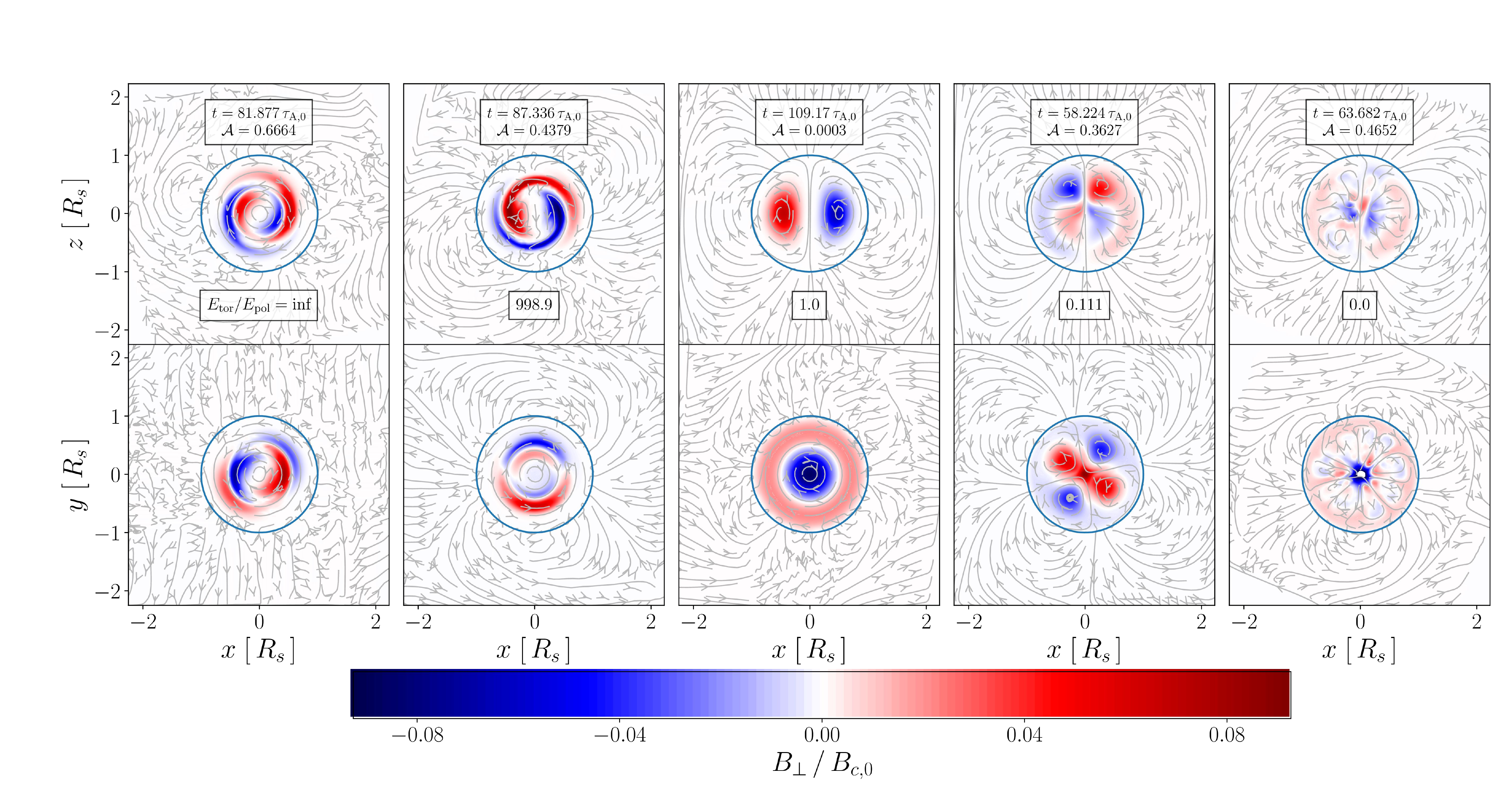}
      \caption{Snapshots of the magnetic field configurations at the indicated times for the initial `Field I'   configuration and different initial values of $E_{\rm tor}/E_{\rm pol}$. The $z$-axis of the simulation box was aligned with the instantaneous magnetic axis $\vec{M}$, so the upper panels show a meridional cut through the star, while the lower panels show an equatorial cut. The color scale corresponds to the strength and direction of the component of the magnetic field perpendicular to the plane shown, while the magnetic field lines correspond to the magnetic field parallel to that plane. 
      The blue circle represents the star's surface. The value of the asymmetry parameter, $\mathcal{A}$, is specified at the top of the upper panels. }
    \label{fig:Bfield_EpEt}
\end{figure*}

\begin{figure*}
    \centering
\includegraphics[width=0.99\textwidth]{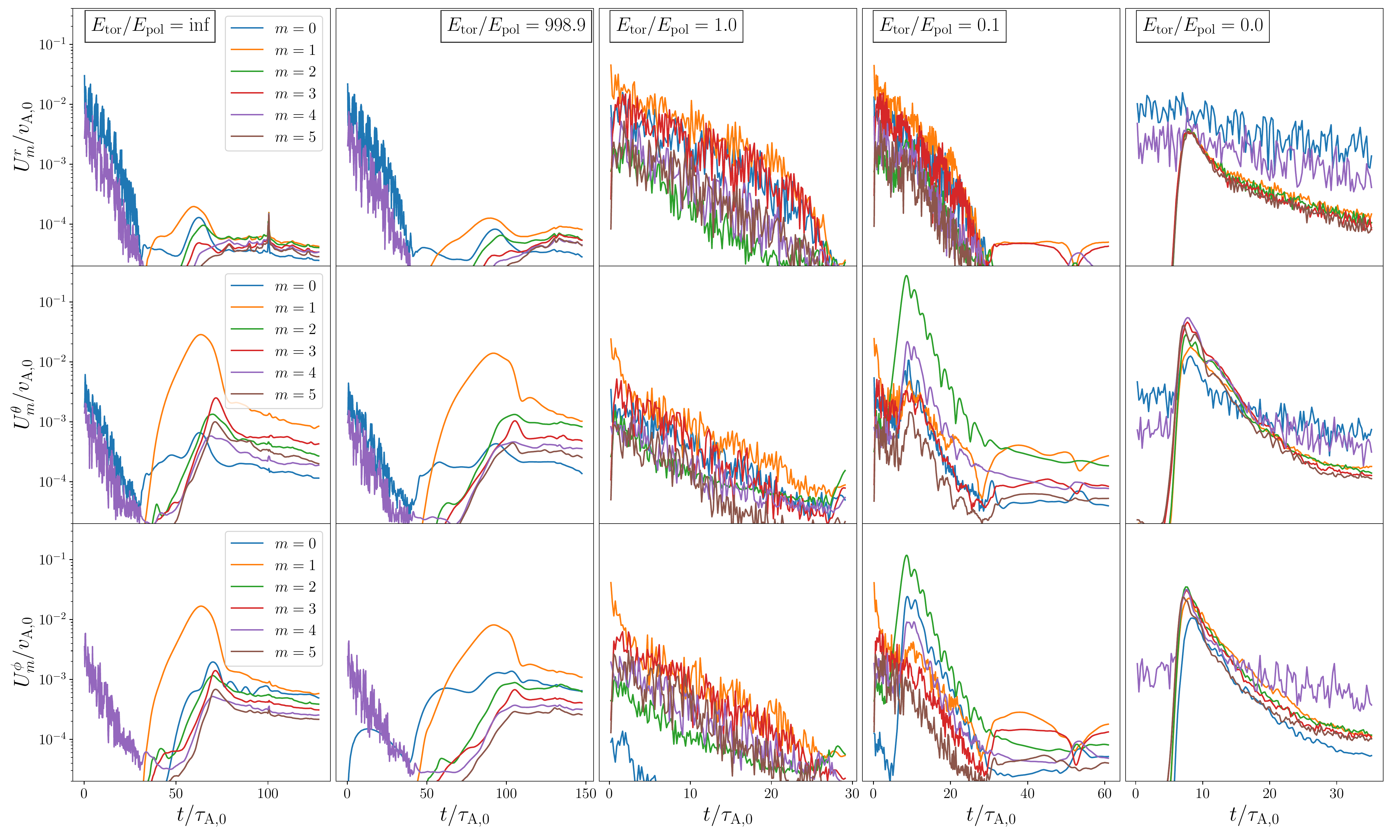}
      \caption{Time evolution of the amplitudes of the azimuthal modes ($m=0$ to $m=5$) for the radial (first row), polar (second row), and azimuthal (third row) components of the velocity field. The different columns correspond to different initial values of $E_{\rm tor}/E_{\rm pol}$.  }
    \label{fig:Modes_EpEt}
\end{figure*}
\begin{figure*}
    \centering
    \subfigure[]{    \includegraphics[width=0.99\columnwidth]{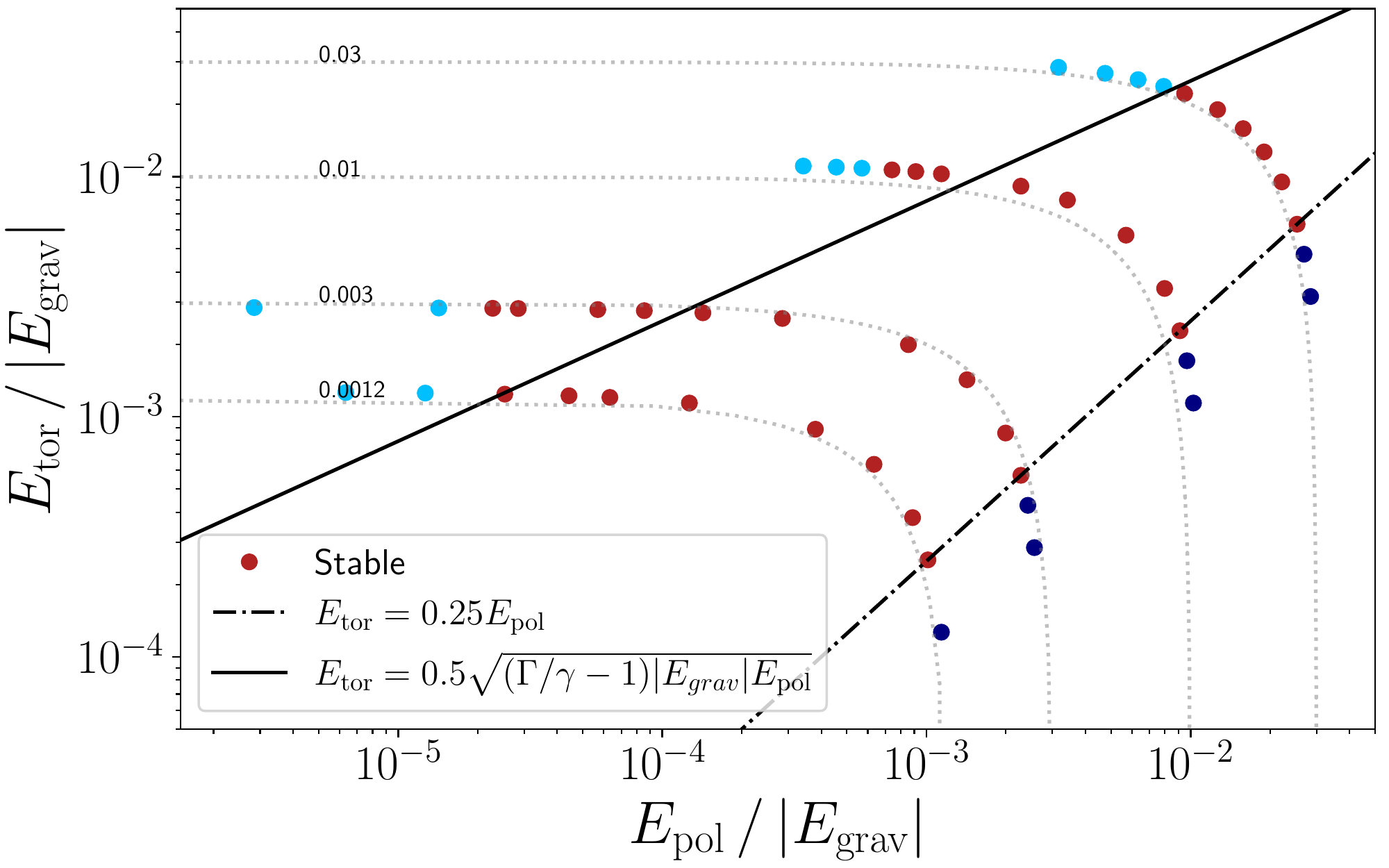}}
     \subfigure[]{
    \includegraphics[width=0.99\columnwidth]{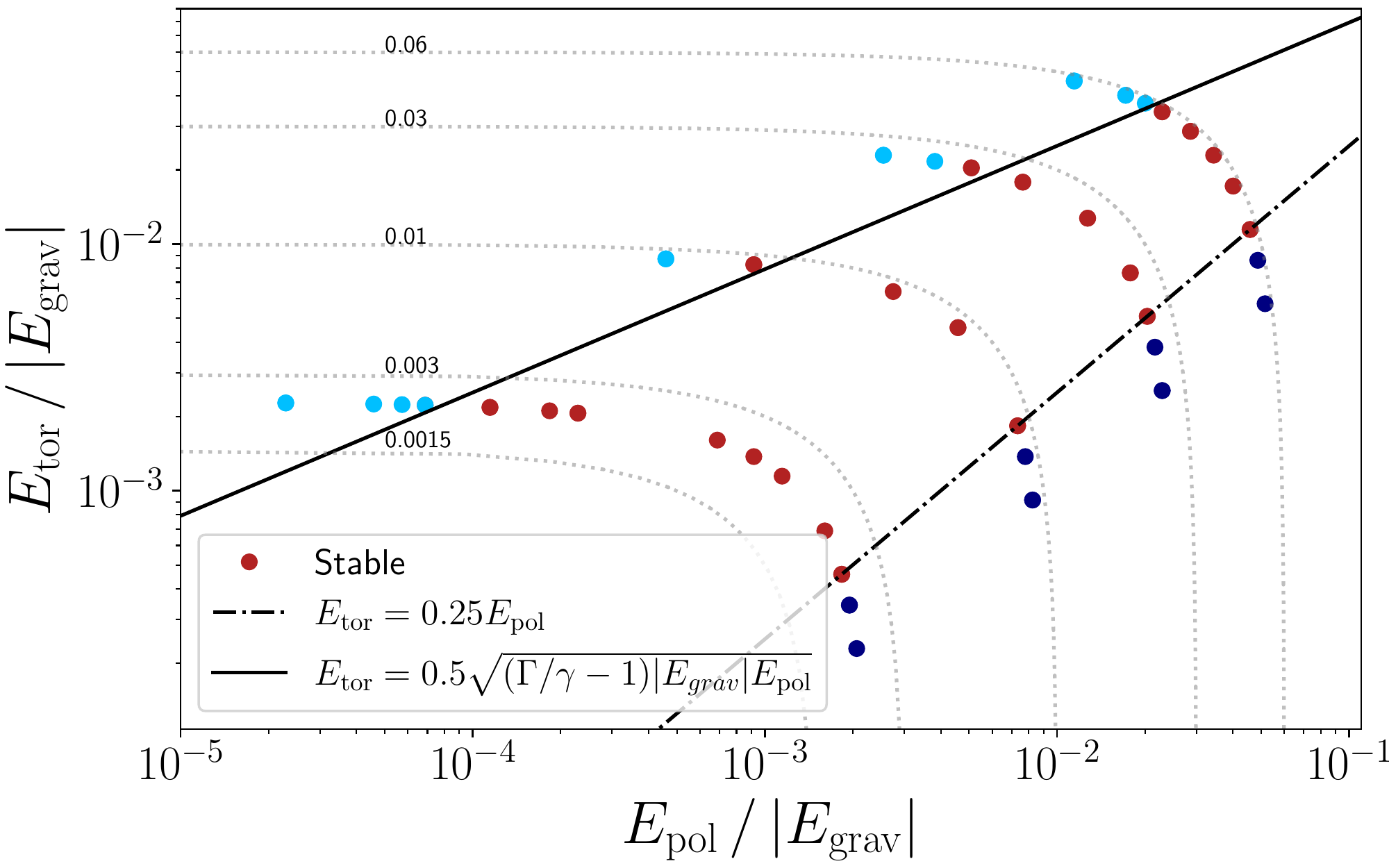}}
    \caption{$E_{\rm tor}/|E_{\rm grav}|$--$E_{\rm pol}/|E_{\rm grav}|$ plane for (a) Field I and (b) Field II magnetic field configurations in stably stratified stellar models with  $\gamma=4/3$. Red points represent stable configurations, while light and dark blue points correspond to unstable configurations.  The dot-dashed and solid lines represent the upper and lower limits in Equation~(\ref{eq:stb_limit}), respectively. Gray dashed lines are lines of constant ratio $E_{\rm mag}/|E_{\rm grav}|$, with values as indicated.}
    \label{fig:Stability_planeI}
\end{figure*}
In our first set of simulations, we constructed a stably stratified star with $\gamma = 4/3$ (and $\Gamma=5/3$), using the Field I magnetic configuration. Figure~\ref{fig:Emag_ArguinE10} shows the time evolution of the magnetic energy (left panel) and the ratio of the kinetic energy to the magnetic energy (right panel) for some representative initial ratios between the energies in the toroidal and poloidal magnetic field components (adjusting the coefficients $a_{\rm pol}$ and $a_{\rm tor}$ of Equation~[\ref{eq:B_axial}]). In all these cases, the initial Alfv\'en travel time is $\tau_{\rm A,0}  = 9.02\,  \tau_{s}$ and the initial magnetic energy is $E_{\rm mag}= 2.8\times 10^{-3} |E_{\rm grav}|$, for which the stability limits of Equation~(\ref{eq:stb_limit}) can be written as $0.25\lesssim E_{\rm tor}/E_{\rm pol}\lesssim 23$. Figure~\ref{fig:Bfield_EpEt} shows snapshots of the magnetic field configuration after several Alfv\'en time scales for some of the instances shown in Figure~\ref{fig:Emag_ArguinE10}. 

For the cases within 
the stability boundaries, namely $E_{\rm tor}/E_{\rm pol}=18.99$, and $1.0$ (and several others not shown here), the magnetic energy decays only very slowly, 
and, for $t\gtrsim\red{0.1}\tau_{\rm A,0}$, the ratio $E_{\rm kin}/E_{\rm mag}$ also drops (almost) monotonically, suggesting a small adjustment to a nearby stable equilibrium. For these cases, the asymmetry parameter remains small: $\mathcal{A}< 10^{-3}$ (see central panel of Figure~\ref{fig:Bfield_EpEt}). 

For the mostly poloidal cases with $E_{\rm tor}/E_{\rm pol}\lesssim 0.2$,  after $\sim 10$ Alfv\'en times, there is a fast decay of the magnetic energy accompanied by a peak in the kinetic energy, signaling an instability. Afterwards,  the magnetic energy becomes almost constant and the kinetic energy drops.  As seen in Figure~\ref{fig:Bfield_EpEt}, for the cases with small, but non-zero $E_{\rm tor}$, we find the magnetic field to evolve to a new non-axisymmetric  magnetic field configuration in a `tennis ball' shape, as in \citet{2009MNRAS.397..763B} and \citet{2015MNRAS.447.1213M}. 

For the other extreme of mostly and purely toroidal fields, $E_{\rm tor}/E_{\rm pol}\gtrsim 150$, the magnetic energy at some time ($t\sim\mathrm{few}\times 10\tau_{\mathrm A,0}$) starts to decay continuously, accompanied by an increase in the kinetic energy, while the field structure becomes increasingly asymmetric. Although the ratio $E_{\rm kin}/E_{\rm mag}$ decreases at late times, $E_{\rm mag}$ keeps decreasing as far as it was possible to run the simulations.\footnote{In Figure 2, the purely toroidal field and the one with a very small poloidal component (which should be stabilizing) seem to become unstable at essentially the same time. We have checked this result with different initial perturbations in the velocity field and found that the former configurations generally become unstable before the latter, confirming our expectations.}  It might be worth noticing that in stably stratified stars the instability sets in substantially later ($t\sim 8-20 \tau_{A,0}$) than in barotropic stars ($t\sim \tau_{A,0}$; see below and \citetalias{paperI}).

As done in \cite{2006A&A...453..687B}, we can separately follow the evolution of each azimuthal mode, $m$, by performing a Fourier decomposition in the $\phi$-direction on each component of the velocity field:

\begin{equation}
\vec{u}(r,\theta,\phi,t) = \sum_{m=0}^{\infty} \vec{u}_m(r,\theta,t) e^{i m\phi}\, ,
\end{equation}
where 
\begin{equation}
    \vec{u}_m(r,\theta,t)={1\over 2\pi}\int_0^{2\pi}d\phi\,\vec{u}(r,\theta,\phi,t)e^{-im\phi}. 
\end{equation}
Figure~\ref{fig:Modes_EpEt} shows the amplitudes, $U_m^i(t)$, of the spherical components ($i=r,\theta,\phi$) of the velocity field (defined as the rms of $|u_m^i(r,\theta,t)|$ over the stellar volume), for modes from $m=0$ to $m=5$ for the same simulations of Figure~\ref{fig:Bfield_EpEt}. In all panels, we see an initial noisy decay of the velocity amplitude. For the stable case ($E_{\rm tor}=E_{\rm pol}$), this behavior continues until essentially no motion occurs. For all the other cases, we later see an increase of (at least) some of the modes, signaling an instability. As expected from the stable stratification of the star (e.~g., \citealt{1973MNRAS.161..365T,2013MNRAS.433.2445A}), the radial component of the velocity field associated with these instabilities is much smaller than the horizontal components ($U_m^r\ll U_m^\theta\sim U_m^\phi$).

 \cite{Goossens+1981} showed that all purely toroidal magnetic fields are dynamically unstable in the relatively small regions that simultaneously satisfy $B_\phi=0$ and $\partial B_\phi^2/\partial \theta>0$, with the fastest growing instabilities being those with $m=1$ azimuthal dependence, which they find to have linear growth rates $\sim 1/t_A$ (or even faster for particular configurations; \citealt{Goossens+1978}). For the purely or mostly toroidal fields (first two columns of Figure~\ref{fig:Modes_EpEt}), we confirm that the most unstable mode is $m=1$, in agreement with the predictions of \citet{1973MNRAS.161..365T} and \cite{Goossens+1981} and the simulation results of \citet{2009MNRAS.397..763B}. Motivated by this result, we also ran simulations with an initially small velocity field $\delta\vec{u}\propto \vec{\nabla}Y_{11}\times \vec{r}$, where $Y_{\ell m}$ is the standard spherical harmonic function \citep[see also][]{2010MNRAS.405..318L}. We found that this makes the instability appear at somewhat earlier times ($\sim 10-20\tau_{A,0}$), and, for simulations with smaller values of the viscosity coefficient, this time decreases to around $9~\tau_{A,0}$ (we note, in any case, that the exponential growth time inferred from the initial increase of the kinetic energy is several times smaller than the time at which the unstable mode appears above the noise level, as read from Figure~\ref{fig:Modes_EpEt}). We checked for the places where the instability starts to grow but they did not coincide with the ones predicted by \cite{Goossens+1981}. Moreover, the instability seen in our simulations appears to be global, involving a motion of the whole star (see Figure~\ref{fig:Emag_ArguinE10}), clearly different from the one studied by \cite{Goossens+1978} and \cite{Goossens+1981}. A general study of the toroidal instability will be pursued in a forthcoming work.

For the mostly poloidal field ($E_{\rm tor}/E_{\rm pol}=0.1$; fourth column of  Figure~\ref{fig:Modes_EpEt}), the most unstable mode is $m=2$, 
while for the purely poloidal field ($E_{\rm tor}/E_{\rm pol}=0$; last column of Figure~\ref{fig:Modes_EpEt}), all azimuthal modes appear to be similarly unstable, in agreement with the linear analysis of \citet{Lander+2011}. 

We ran  similar simulations (always for the `Field I' magnetic field configuration) changing the initial strength of the magnetic field to $E_{\rm mag}/ |E_{\rm grav}| = 1.2\times 10^{-3}$, $0.011$, and $0.025$, which correspond to $\tau_{\rm A,0}/\tau_{s} =13.5$,  $4.5$, and $3.0$, respectively. Figure~\ref{fig:Stability_planeI}(a) summarizes the results in the $E_{\rm tor}/|E_{\rm grav}|$--$E_{\rm pol}/|E_{\rm grav}|$ plane. The types of behaviour found are as described in the previous paragraphs, which allowed us to distinguish quite clearly which of the initial magnetic field configurations were stable and which were not. We can see that, for this set of simulations, the lower boundary of the stable region defined by Equation~(\ref{eq:stb_limit}) fairly accurately represents our numerical results, whereas the upper boundary only gives a very rough approximation. 
 
The same study was done for the Field II magnetic field configurations, and the results are summarized in Figure~\ref{fig:Stability_planeI}(b). We ran simulations with magnetic energies $E_{\rm mag}/|E_{\rm grav}| = 2.2\times 10^{-3}$, $8.9\times 10^{-3}$, $0.023$, and $0.053$. In all these cases, Equation~(\ref{eq:stb_limit}) describes both boundaries of the stability region reasonably well.

Unfortunately, simulations with higher ratios $E_{\rm mag}/|E_{\rm grav}|$ suffer from numerical instabilities that do not allow us to follow their evolution and check the stability of very strong magnetic fields. 

 \subsection{Dependence on the star's stable stratification}\label{sec:stb_stra}

\begin{figure}
    \centering
     \includegraphics[width=\columnwidth]{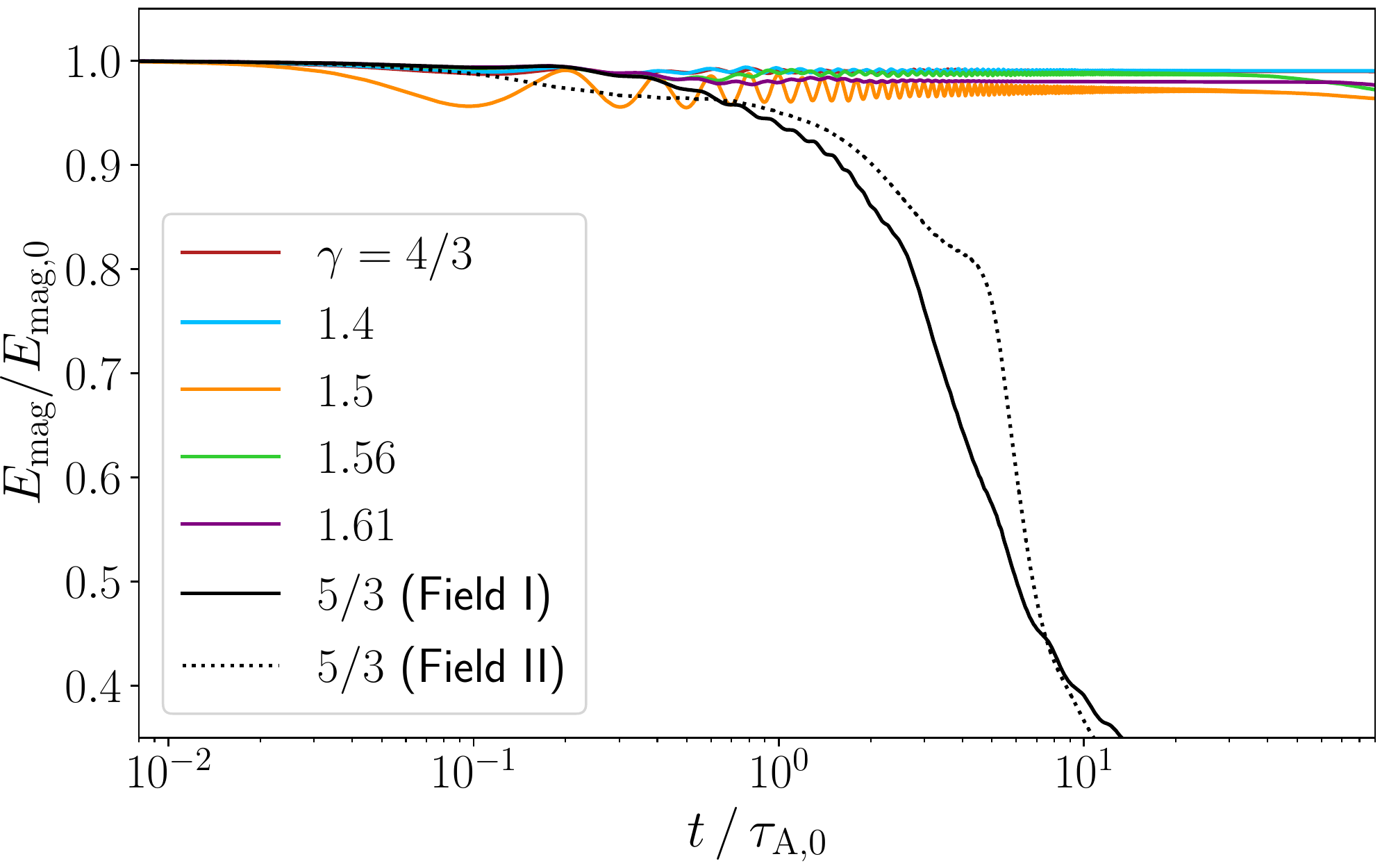}
  
    \caption{Evolution of the total magnetic energy for different specific entropy gradients. 
    The initial magnetic field configuration corresponds to Field~I with $E_{\rm tor}=E_{\rm pol}$, and $E_{\rm mag}= 2.8\times 10^{-3}~|E_{\rm grav}|$. We also show for reference the case of Field II for $\gamma=5/3$. }
    \label{fig:Emag_poly}
\end{figure}

Finally, we explored the accuracy of Equation~(\ref{eq:stb_limit}) for different degrees of stable stratification of the star. Figure~\ref{fig:Emag_poly} shows the evolution of the magnetic energy for $\Gamma=5/3$ and an initial Field I configuration with $E_{\rm tor}=E_{\rm pol}$, $\tau_{\rm A,0}\approx 9~\tau_s$, and $E_{\rm mag} \approx 2.8\times 10^{-3}~|E_{\rm grav}|$  for different values of $\gamma$.  For the cases when $\Gamma>\gamma$, even for the very slightly non-barotropic one ($\gamma=1.61$), the magnetic energy  decays on a much longer timescale than the Alfv\'en timescale, indicating stability, consistent with Equation~(\ref{eq:stb_limit}). 

\begin{figure}
    \centering
     \includegraphics[width=\columnwidth]{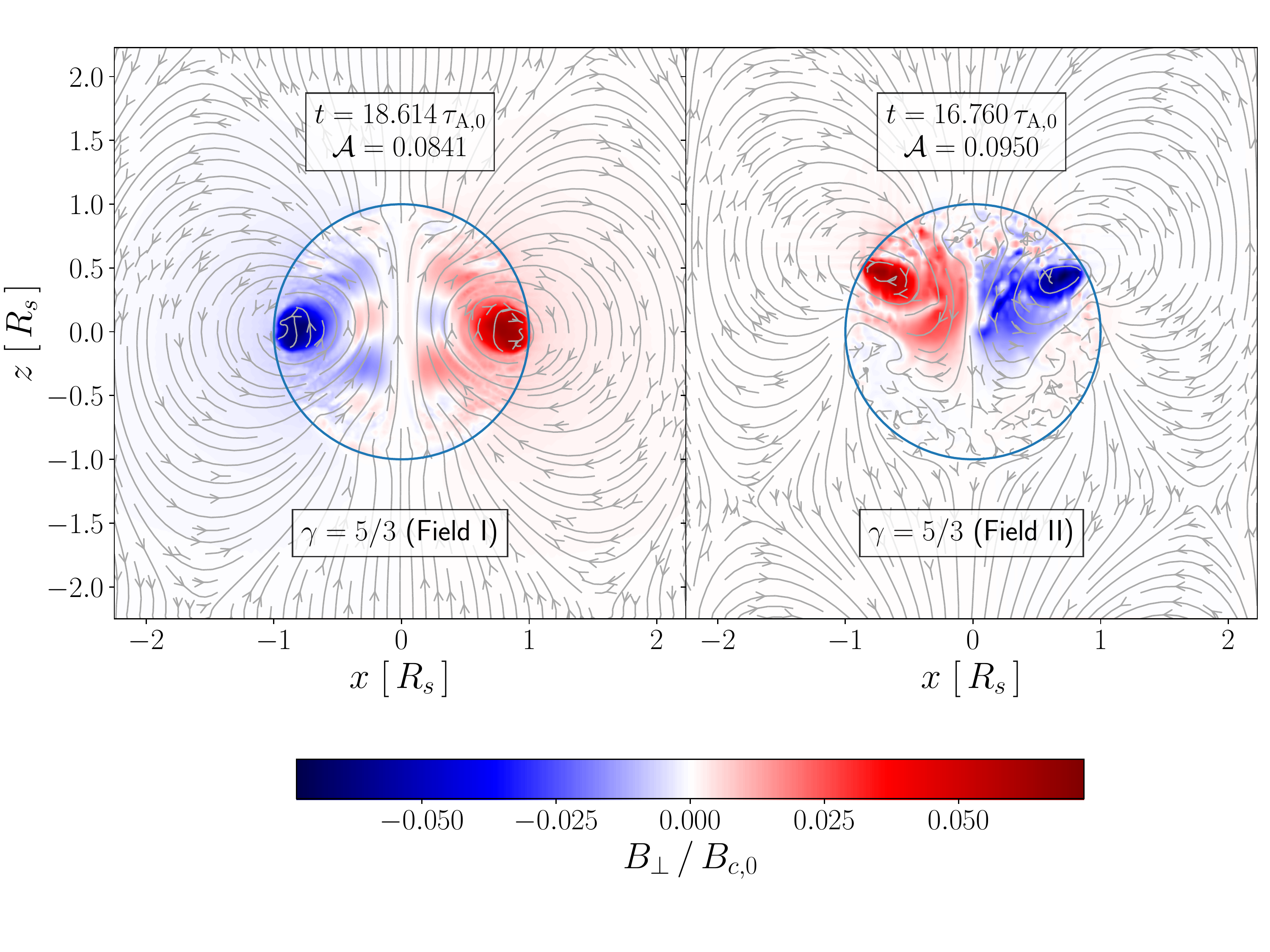}
  
    \caption{Snapshots of the magnetic field for a barotropic star ($\Gamma=\gamma$) at the indicated times for the initial `Field I'(left panel) and `Field II' (right panel)  configurations with $E_{\rm tor}=E_{\rm pol}$, and $E_{\rm mag}= 2.8\times 10^{-3}~|E_{\rm grav}|$.  The colour scale and field lines are the same as in Fig.~\ref{fig:Bfield_EpEt}.
    }
    \label{fig:Bfield_barotropic}
\end{figure}

Figure~\ref{fig:Emag_poly} also shows the evolution of the magnetic energy for Fields I and II when $\gamma = \Gamma = 5/3$ (barotropic case).  
In both cases, the magnetic energy quickly drops to half its initial value at about $10~\tau_{A,0}$. As seen in Figure~\ref{fig:Bfield_barotropic}, by this time the magnetic field has moved to the star's surface,  roughly preserving its axial symmetry ($\mathcal{A}<0.1$). Unfortunately, beyond this point, the numerical inaccuracies increase considerably (the energy balance equation is no longer satisfied with good accuracy) and can no longer be neglected, making the simulation unreliable. We expect that, due to the magnetic buoyancy, a large portion of the magnetic field will get expelled from the star and eventually be dissipated by the magnetic diffusivity of the atmosphere. We believe that these numerical problems originate near the surface of the star, where the transition to the uniform temperature atmosphere happens. They do not appear in the stably stratified cases ($\Gamma>\gamma$), even in the slightly stratified ones, probably because the magnetic field does not move to the star’s surface, as it does in the barotropic case.  Moreover, we recall that, due to our numerical setup (see Section~\ref{sec:setup}), the star is not completely barotropic, but has a thin stably stratified surface layer that could be preventing the magnetic field from being completely expelled.  

\begin{figure*}
    \centering
     \includegraphics[width=\textwidth]{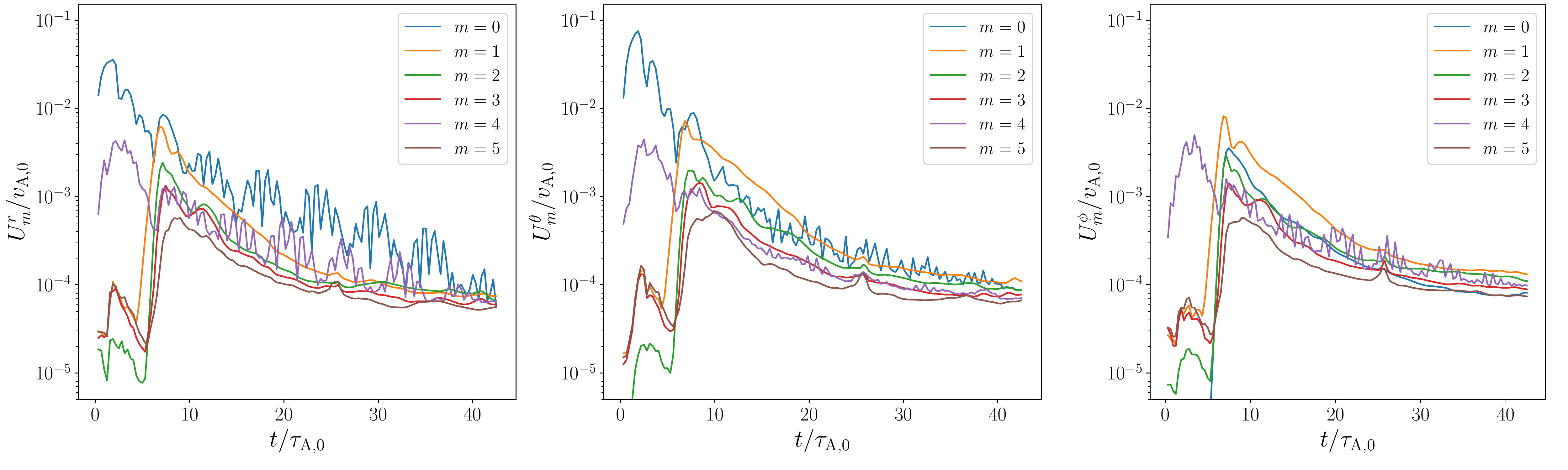}
  
    \caption{ Same as Figure~\ref{fig:Modes_EpEt} for a barotropic star ($\Gamma=\gamma=5/3$) with the initial Field I configuration with $E_{\rm tor}/E_{\rm pol}=\infty$. }
    \label{fig:Modes_barotropic}
\end{figure*}

 Figure~\ref{fig:Modes_barotropic} shows the time evolution of the amplitudes of the azimuthal modes in a  barotropic star with the initial Field I configuration with $E_{\rm tor}/E_{\rm pol}=\infty$. In contrast with the stably stratified case (first column of Figure~\ref{fig:Modes_EpEt}), a clear instability appears at $\sim \tau_{A,0}$, with all velocity components of  similar amplitude, $U_1^r\sim U_1^\theta\sim U_1^\phi$. 
 
\begin{figure*}
    \centering
   \subfigure[]{ \includegraphics[width=0.48\textwidth]{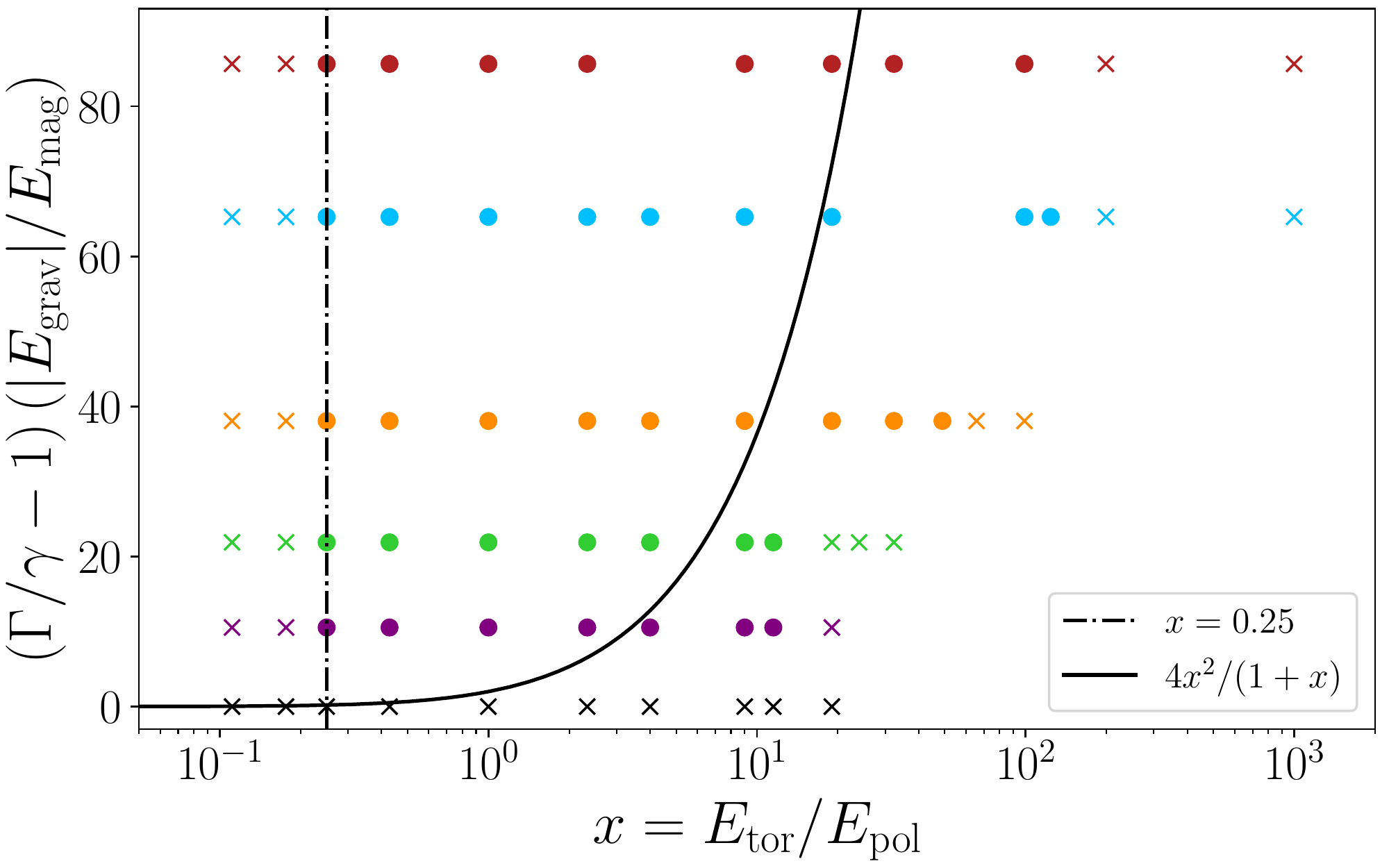}}
    \subfigure[]{\includegraphics[width=0.48\textwidth]{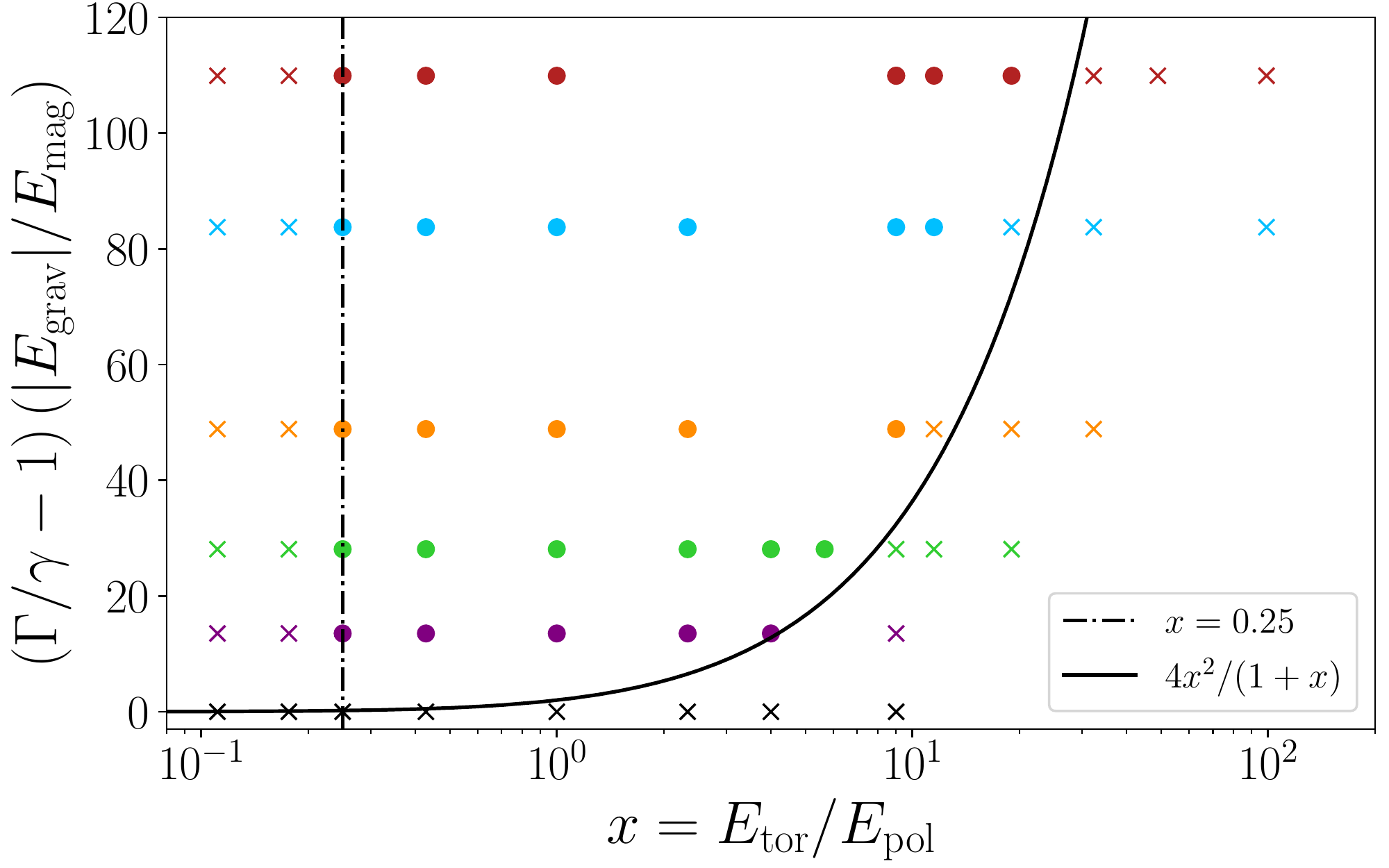}}
    
    \caption{Plane of $x\equiv (E_{\rm tor}/E_{\rm pol})$ vs. $(\Gamma/\gamma-1)(|E_{\rm grav}|/E_{\rm mag})$ for (a) Field I and (b) Field II 
    and different values of the polytropic index $\gamma=1+1/n$, always with adiabatic index $\Gamma=5/3$ and $E_{\rm mag}=2.8\times 10^{-3}~|E_{\rm grav}|$. Filled circles represent configurations found numerically to be stable, while crosses signal unstable configurations. The dash-dotted and solid lines correspond to the lower and upper limit, respectively, for stability established in Equation~(\ref{eq:stb_limit}).
    }
    \label{fig:Stb_poly}
\end{figure*}

We ran additional simulations for different values of $E_{\rm tor}/E_{\rm pol}$.  Figure~\ref{fig:Stb_poly} summarizes the results of these simulations on the $(E_{\rm tor}/E_{\rm pol})$ vs. $(\Gamma/\gamma-1)(|E_{\rm grav}|/E_{\rm mag})$ plane for the Field I and Field II magnetic field configurations, showing Equation~(\ref{eq:stb_limit})  to be more accurate for the latter.

\section{Discussion and conclusions}\label{sec:conclusion}

We did a parameter-space study to explore the validity of the proposed stability condition for axially symmetric magnetic fields in stars given by Equation~(\ref{eq:stb_limit}). It is worth noticing that we have used the Cowling approximation  to speed up the simulations, which  has been shown to lead to an overestimation of the system's stability \citep{1973MNRAS.161..365T}. We verified the stability of axially symmetric magnetic fields even in very slightly non-barotropic stars ($\Gamma=5/3$ and $\gamma = 1.61$), but we did not find any stable magnetic field configurations in  barotropic stars ($\Gamma=\gamma=5/3$). These simulations confirm previous suggestions that the star's stable stratification is a crucial ingredient for the stability of magnetic fields inside stars \citep{2009A&A...499..557R,2015MNRAS.447.1213M}. Of course, neither the previous simulations nor those reported here can prove that there are no stable equilibria in barotropic stars. If there were any, they should satisfy the Grad-Shafranov equation \citep{grad1958,1958JETP....6..545S}. As mentioned in the Introduction, when the magnetic field is not totally confined inside the star, the numerical solutions of this equation found by various authors all satisfy $E_{\rm tor}<0.1E_{\rm pol}$, so they lie below the lower limit of the stability region defined by equation~\ref{eq:stb_limit}. We saw in the simulations that, in stably stratified stars, such configurations evolve into non-axisymmetric stable equilibria, while for barotropic stars the magnetic field moves to the star's surface, where our numerical setup becomes unreliable, so we cannot follow the further evolution of the magnetic field.

We verified that the lower limit for $E_{\rm tor}/E_{\rm pol}$ in Equation~(\ref{eq:stb_limit}) is quite robust,  
a toroidal field with $E_{\rm tor}\gtrsim 0.25 E_{\rm pol}$ is needed to stabilize the poloidal field, independent of the magnetic field configuration or the star's stratification, in agreement with the earlier analytical work of \cite{1973MNRAS.162..339W, 1973MNRAS.163...77M} and the simulations of \citet{2009MNRAS.397..763B}.

On the other hand, the upper bound for $E_{\rm tor}/E_{\rm pol}$ in Equation~(\ref{eq:stb_limit}) appears to be somewhat less robust, as we found some numerically stable configurations beyond it.
In addition to the star's stable stratification and gravitational energy (considered in Equation [\ref{eq:stb_limit}]), this limit appears to depend on the geometry of the magnetic field.
Additionally, many works have established that rotation reduces the growth rate of the Tayler instability \cite[see, e.g.,][]{1985MNRAS.216..139P, 1999A&A...349..189S} and might have a similar effect on others instabilities.

Extrapolating Equation~(\ref{eq:stb_limit}) to very strong magnetic fields (where its applicability is questionable) implies an upper bound for the total magnetic energy: 
\begin{equation} \label{eq:upperlimit}
    E_{\rm mag}\lesssim 5(\Gamma/\gamma-1)|E_{\rm grav}|,
\end{equation}
with comparable fractions in the poloidal and toroidal components, $E_{\rm pol}^{\rm max}\sim 4E_{\rm tor}^{\rm max}$. For the Ap/Bp stars, the fluid can be approximated as a classical monatomic ideal gas with $\Gamma=5/3$ and $\gamma=4/3$ \citep{2003ApJ...586..480M}, so $E_{\rm mag}\lesssim {5\over 4}|E_{\rm grav}|$, very similar to the rigorous upper limit implied by the virial theorem, $E_{\rm mag}\leq |E_{\rm grav}|$. From this condition, the maximum allowed value for $B_{\rm rms}$ in Ap/Bp stars is $\sim 5\times10^7$~G. For degenerate stars, since $\Gamma/\gamma-1\ll 1$, Equation~(\ref{eq:upperlimit}) is more constraining than the virial theorem. In white dwarfs, $\Gamma/\gamma-1\sim T_7/500$ with $T=T_7\times 10^7$~K \citep{2009A&A...499..557R}, implying $B_{\rm rms, max}\sim 10^{11} T_7^{1/2}$~G, whereas in neutron stars $\Gamma/\gamma-1\sim Y$ ($Y$ is the proton fraction; \citealt{1992ApJ...395..240R}), so  $B_{\rm rms, max}\sim 3\times 10^{17}\left(Y/0.1\right)^{1/2}$~G. The observed surface magnetic fields in all these stars are far below these limits.

We note that our application of the present results to neutron stars assumes that they are composed of a ``normal'' fluid, that is, ignoring the formation of a solid crust and the transition of protons and neutrons to a superconducting and superfluid state, respectively. Most of the crust (its inner regions) freezes after a few years \citep{kha15}; while protons are expected to become superconducting in a few months (or even faster, if direct URCA processes are allowed; \citealt{gyp01}), and neutrons to become superfluid in a few hundred years \citep{syhhp11}. The crust has a stabilizing effect on the magnetic field, which can be important unless the field strength is extremely high (e.~g., \citealt{Lander+2019} and references therein). The transition of the protons to a superconducting state causes the confinement of the magnetic field into magnetic flux tubes (for type-II superconductors) or other small-scale structures (for type-I superconductors), increasing the magnetic energy and thus the Lorentz force for a given magnetic flux \citep{Easson+1977}. In the absence of rotation, the superfluid transition of the neutrons decouples them from the charged particles, and thus from the magnetic field \citep{Gusakov+2020}. In low-density regions of the core, where protons and electrons are the only charged particles, they act as a barotropic fluid, which cannot by itself stabilize the magnetic field. However, at higher densities, there are additional charged particles, such as muons, whose relative abundance is a function of density, making the charged-particle fluid stably stratified. In this case, the magnetic field could be stabilized by this stably stratified fluid, making the results of this paper at least qualitatively applicable. The magnetic field should be non-zero also in the barotropic region, but there it will be constrained to satisfy the Grad-Shafranov equation. In a rotating neutron star, the quantized neutron vortices interact with the magnetic field, providing a coupling between neutrons and charged particles whose effect remains to be understood in detail \citep{Dommes+2021}.

\section*{Acknowledgements}

The authors are grateful to J. Braithwaite, F. Castillo, N. Moraga, D. Ofengeim, and H. Spruit for useful discussions, and to the Referee for stimulating comments and suggestions. This work was supported by FONDECYT projects 3190172 (L.B.), 1201582 (A.R.), and 1190703 (J.A.V.), and the Center for Astrophysics and Associated Technologies (CATA; ANID Basal project FB210003). J.A.V. thanks for the support of CEDENNA under ANID/CONICYT grant FB0807. M.E.G. acknowledges support from the Russian Science Foundation [Grant No.\ 22-12-00048]. The calculations presented in this work were performed on the Geryon computer at the UC Center for Astro-Engineering, part of the BASAL PFB-06 and AFB-170002 projects, which received additional funding from QUIMAL 130008 and Fondequip AIC-57 for upgrades.

\section*{Data Availability}

The data underlying this article will be shared on reasonable request to the corresponding author.



\bibliographystyle{mnras}
\bibliography{biblio}



\appendix

\section{Equations }\label{ap:equations}  

The MHD equations solved in the simulations presented in this paper are:
\begin{eqnarray}\label{eq:MHDEqs_a}
   \frac{\partial ({\rm ln}\,\rho)}{\partial t} &=& -\vec{\nabla}\cdot\vec{u} - \vec{u} \cdot \vec{\nabla}({\rm ln}\,\rho)  \\
   \label{eq:MHDEqs_b}
   \frac{\partial \vec{u}}{\partial t} &=&  -\vec{u}\cdot\vec{\nabla}\vec{u} - \frac{\vec{\nabla}p}{\rho}-\vec{\nabla}\Phi + \frac{\vec{j}\times\vec{B}}{\rho} +   \vec{f}_{\rm visc} \\
 \label{eq:MHDEqs_c}
   \frac{\partial\vec{A}}{\partial t} &=& \vec{u}\times\vec{B} -\eta \mu_0\vec{j}\\
   \label{eq:MHDEqs_d}
    \frac{\partial s}{\partial t} &=& -\vec{u}\cdot \vec{\nabla} s + \frac{\eta\mu_0|\vec{j}|^2}{\rho T} + \frac{2\nu\mathbf{S}^2}{T}\, ,
\end{eqnarray}
where $\vec{A}$ is the magnetic vector potential, $\vec{B}=\vec{\nabla}\times\vec{A}$ is  the magnetic  field, $\vec{j}=\mu_0^{-1}\vec{\nabla}\times\vec{B}$ is the current density, and $\mu_0$  the magnetic vacuum permeability.  Here, $\Phi$ is the gravitational potential, $\rho$ is the fluid mass density, $p$ is its pressure, $\vec{u}$  is the fluid velocity, and $s$ is the entropy per unit mass.  The viscous force is given by
\begin{equation}\label{eq:visc_force}
 \vec{f}_{\rm visc}=\rho^{-1}\vec{\nabla}\cdot\left(2\rho \nu{\mathbf{S}} \right) \, ,
\end{equation}
with   $\nu$ as the kinematic viscosity, and $\mathbf{S}$ as the rate-of-shear tensor whose components are
\begin{equation}\label{eq:shear_tensor}
S_{ij}= \frac{1}{2}\left[\frac{\partial u_j}{\partial x_i}+ \frac{\partial u_i}{\partial x_j}-\frac{2}{3}\delta_{ij}(\vec{\nabla}\cdot\vec{u})\right]\, .
\end{equation}
We assume an ideal equation of state:
\begin{equation}\label{eq:pres_def}
    p(\rho,s)=(\mathcal{R}/\mu) \rho T(\rho,s) \, ,
\end{equation}
with,
\begin{equation}\label{eq:tem_def}
    T(\rho,s) = T_c \left(\frac{\rho}{\rho_c} \right) ^{\Gamma-1} \exp{[(s-s_c)/c_V]}\, ,
\end{equation}
where $s_c$, $\rho_c$ and $T_c$ are the entropy, density, and temperature at the center of the star, respectively.



\bsp	
\label{lastpage}
\end{document}